\documentstyle[aps,multicol,epsfig,epsf,psfrag]{revtex}

\draft
\newcommand{\beq}{\begin{equation}}
\newcommand{\eeq}{\end{equation}}
\newcommand{\bdis}{\begin{displaymath}}
\newcommand{\edis}{\end{displaymath}}
\newcommand{\bea}{\begin{eqnarray}}
\newcommand{\eea}{\end{eqnarray}}
\newcommand{\barr}{\begin{array}}
\newcommand{\earr}{\end{array}}

\begin{document}

\title{Scaling laws of creep rupture of fiber bundles}

\author{Ferenc
  Kun$^{1}$\footnote{Electronic
address:feri@dtp.atomki.hu}, Raul
Cruz Hidalgo$^{2}$, Hans. J. Herrmann$^{2}$, and K\'aroly F.\ P\'al$^{3}$} 

\address{$^1$Department of Theoretical Physics, University of Debrecen, \\ 
P.O.Box: 5, H-4010 Debrecen, Hungary \\
$^2$Institute for Computational Physics, University of
Stuttgart,    
Pfaffenwaldring 27, 70569 Stuttgart, Germany\\
${}^3$Institute of Nuclear Research (ATOMKI), P.\ O.\ Box 51, H-4001
Debrecen, Hungary
}

\date{\today}
\maketitle
\begin{abstract} 
We study the creep rupture of fiber composites in
the framework of fiber bundle models.
Two novel fiber bundle models are introduced based on different 
microscopic mechanisms responsible for the macroscopic creep 
behavior.
Analytical and numerical calculations show that above a critical
load the deformation of the creeping system monotonically
increases in time resulting in global failure at a finite time $t_f$,
while below the critical load the system suffers only partial failure
and the deformation tends to a constant value giving rise to an
infinite lifetime.  
It is found 
that approaching the critical load from below and above the creeping
system is characterized by universal power laws when the fibers have
long range interaction.  
The lifetime of the composite above the
critical point has a universal dependence on the system size. 
\end{abstract}


\begin{multicols}{2}
\narrowtext

\section{introduction} 
Under high steady stresses fiber composites may undergo time dependent
deformation resulting in failure called creep rupture which limits
their lifetime, and hence, has a high impact on the applicability of
these materials in construction elements. Both
natural fiber composites like wood
\cite{laufenberg,pu,acoust,gerhards} and 
various types of fiber reinforced composites
\cite{chiao,otani,farq,phoenix,mcmeek,curtin} 
show creep 
rupture phenomena, which have attracted continuous theoretical
and experimental interest over the past years. The underlying
microscopic failure mechanism of creep   
rupture is very complex depending on several characteristics of
the specific types of materials, and is far from being well
understood. Theoretical studies encounter various challenges: on the one
hand, applications of fiber composites require  
the development of analytical and numerical models which are able to
predict the damage histories of loaded composites in terms of the
specific parameters of constituents. On the other hand, creep
rupture, similarly to other rupture phenomena, presents a very
interesting problem for statistical physics, it is still an open
problem to embed  creep rupture into the general framework of
statistical physics and to understand the analogy between rupture
phenomena and phase transitions. 

Creep failure tests are usually
performed under uniaxial tensile loading when the
specimen is subjected either 
to a constant load $\sigma_{\rm o}$ or to an increasing load
(ramp-loading) and the time evolution of the 
damage process is monitored by recording the strain $\varepsilon$ of
the specimen and
the acoustic signals emitted by microscopic failure events
\cite{laufenberg,pu,acoust,gerhards,chiao,otani,farq,phoenix,mcmeek,curtin}.
In the present paper we study the creep rupture of fiber composites in 
the framework of fiber bundle models
\cite{coleman,daniels,hansen,phoenix1,leath,zhang,delapl,sornette1,sornette2,sornette3,kun1,kun3,yamir,menezes}
focusing on general characteristics of creep rupture.  Our main goal
is to reveal the universal aspects of a 
creeping system which do not depend on specific
material properties. We will derive
scaling laws which emerge when
approaching the failure stress both from below and above, furthermore,
the finite size scaling of the time to failure is studied.  Beyond the
theoretical understanding, exploring universal features of creep helps
to evaluate experimental data and can serve as a guide to extract the
relevant information invariant under the variation of specific material 
properties. 

 We consider only the case of global load sharing (GLS) for the
redistribution of load following fiber failure
\cite{hansen,sornette1,sornette2,sornette3,kun1,kun3,yamir}. For several
types of 
materials GLS provides an adequate approach, and it has the advantage
that many of the important GLS results can be obtained in closed
analytic forms. During creep at 
the critical point and above it, stress localization occurs which
can be captured by localized load sharing in fiber bundle models
\cite{hansen,delapl,kun1} but
this is beyond the scope of our investigations.  
The present study is focused on analytic modeling in
the framework of GLS. We also develop efficient simulation techniques
for testing the analytic results, furthermore, to study finite size
system. Since similar studies with local load sharing require large scale 
computer simulations and a large amount of data processing it will be
presented in a forthcoming publication. 

\section{models}
In order to work out a
theoretical description of creep failure we introduce two novel fiber
bundle models improving the classical fiber bundle models
\cite{coleman,daniels} which have proven very successful in the  
study of fracture of disordered materials
\cite{hansen,phoenix1,leath,zhang,delapl,sornette1,sornette2,sornette3,yamir}.
The two models differ in the microscopic mechanisms assumed to be
responsible for 
the macroscopic creep behavior: $(i)$ in the first approach the fibers 
themselves are viscoelastic and they break when their
deformation exceeds a stochastically distributed threshold value
(strain controlled breaking), $(ii)$ in the second model
the fibers are linearly elastic until they break stochastically in a
stress controlled way,
however, after breaking their relaxation is not instantaneous but  
the sliding of the broken fiber with respect to the matrix material,
and the yielding or creeping of the matrix 
introduces an intrinsic time scale for the relaxation.

The mathematical formalism of the first model (viscoelastic bundle) is 
more simple, hence, the steps of analysis will be presented in details 
for this case. For the second model the corresponding results will be
summarized. 
\subsection{Viscoelastic fiber bundle}

\subsubsection{Analytic model}
\begin{figure} 
\begin{center}
\epsfig{bbllx=0,bblly=-20,bburx=250,bbury=210, file=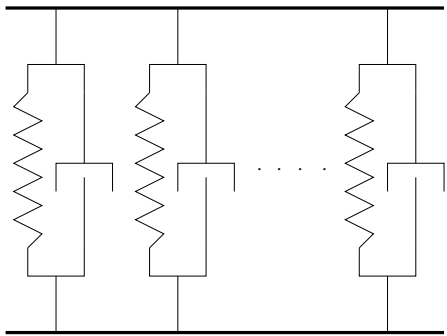, width=6cm}
\caption{The viscoelastic fiber bundle: intact fibers are modeled by
  Kelvin-Voigt elements. After fiber breaking the corresponding
  element is removed from the model.}
\label{fig:kelvin}
\end{center}
\end{figure} 
Our model consists of $N$ parallel fibers having
viscoelastic constitutive behavior. 
For simplicity, the pure viscoelastic behavior of fibers is
modeled by a Kelvin-Voigt element which consists of a spring and a
dashpot in parallel (see Fig.\ \ref{fig:kelvin}) 
and results in the constitutive equation
\begin{eqnarray}
  \label{eq:visco}
  \sigma_{\rm o} = \beta \dot{\varepsilon} + E\varepsilon, 
\end{eqnarray}
where $\beta$ denotes the damping coefficient, and  $E$ the Young
modulus of fibers, respectively.  
 Eq.\ (\ref{eq:visco}) provides
the time dependent deformation $\varepsilon(t)$ of a fiber at a
fixed external load  $\sigma_{\rm o}$
\begin{eqnarray}
  \label{solve}
  \varepsilon(t) = \frac{\sigma_o}{E}\left[1-e^{-Et/\beta} \right] +
  \varepsilon_{\rm o} e^{-Et/\beta},
\end{eqnarray}
where $\varepsilon_o$ denotes the initial strain at $t=0$. It can be
seen that  $\varepsilon(t)$ converges to $\sigma_{\rm o}/E$ for $t 
\to \infty$, which implies that the asymptotic strain fulfills Hook's
law. To incorporate breaking in the model we introduce a strain controlled
failure criterion for fibers: a fiber fails 
during the time evolution of the system if
its strain exceeds a damage threshold $\varepsilon_d$, which is an
independent identically distributed random variable of fibers with
probability density 
$p(\varepsilon_d)$ and cumulative distribution ${\displaystyle 
P(\varepsilon_d) = \int_0^{\varepsilon_d} p(x) dx}$. 
Due to the
validity of Hook's law for the asymptotic strain values, the 
formulation of the failure criterion in terms of strain instead of
stress implies that under a certain steady load the same amount of
damage occurs  as in the case of stress controlled failure, however,
the breaking of fibers is not instantaneous but distributed over
time. 
When a fiber fails its load has to be redistributed to the
intact fibers. Assuming global load sharing, the time evolution of the
system under a steady external load $\sigma_o$ is finally described by
the equation  
\begin{eqnarray}
  \label{eq:eom}
  \frac{\sigma_{\rm o}}{1-P(\varepsilon)} = \beta \dot{\varepsilon}
  +E\varepsilon, 
\end{eqnarray}
where the viscoelastic behavior of fibers is coupled to the failure 
of fibers in a global load sharing framework \cite{kun4}. 

For the behavior of the solutions of Eq.\ (\ref{eq:eom}) two distinct
regimes can be distinguished depending on the value of the external
load $\sigma_{\rm o}$: When $\sigma_{\rm o}$ is below a critical value
$\sigma_{\rm c}$ Eq.\ (\ref{eq:eom}) has a stationary solution
$\varepsilon_s$, which can be obtained by setting
$\dot{\varepsilon}=0$ in Eq.\ (\ref{eq:eom})
\begin{eqnarray}
  \label{eq:stationary}
  \sigma_{\rm o} = E\varepsilon_s[1-P(\varepsilon_s)].
\end{eqnarray}
It means that until this equation can be solved for $\varepsilon_s$ 
at a given external load $\sigma_{\rm o}$, the solution  $\varepsilon(t)$ of
Eq.\ (\ref{eq:eom}) converges to 
$\varepsilon_s$  when $t\to \infty$, and no macroscopic failure
occurs. However, when 
$\sigma_{\rm o}$  exceeds the critical value $\sigma_{c}$ no stationary
solution exists, furthermore, $\dot{\varepsilon}$ remains always
positive, which implies that for $\sigma > \sigma_{c}$ the strain of the
system   $\varepsilon(t)$ monotonically increases until the 
system fails globally at a time $t_f$ \cite{kun4}. 

In the regime $\sigma_{\rm o} \leq \sigma_{\rm c}$ Eq.\
(\ref{eq:stationary}) also provides the asymptotic constitutive
behavior of the 
fiber bundle which can be measured by controlling the external load
$\sigma_{\rm o}$ and letting the system relax to $\varepsilon_s$. It follows 
from the above argument that the critical value of the load
$\sigma_{c}$ is the static fracture strength of the bundle which can be
determined from Eq.\ (\ref{eq:stationary}) as 
 $ \sigma_{c} = E\varepsilon_{c}[1-P(\varepsilon_{c})]$,
where $\varepsilon_{c}$ is the solution of the equation
$\displaystyle{\left. d\sigma_{\rm o}/d\varepsilon_s
  \right|_{\varepsilon_{\rm c}} = 0}$ 
\cite{sornette1}.    
Since $\sigma_{\rm o}(\varepsilon_{\rm s})$ has a maximum of the value
$\sigma_{\rm c}$ at 
$\varepsilon_{\rm c}$, in the vicinity of $\varepsilon_{\rm c}$ it can be
approximated as  
\begin{eqnarray}
  \label{eq:series}
  \sigma_{\rm o} \approx \sigma_{\rm c} -A(\varepsilon_{\rm
    c}-\varepsilon_s)^2, 
\end{eqnarray}
where the multiplication factor $A$ depends on the probability
distribution $P$.
A complete description of the system can be obtained by solving the 
differential equation Eq.\ (\ref{eq:eom}). After separation of
variables 
the integral arises 
\begin{eqnarray}
  \label{eq:integ}
t =  \beta \int d\varepsilon 
  \frac{1-P(\varepsilon)}{\sigma_{\rm o}-E\varepsilon \left[
    1- P(\varepsilon)\right]} + C,
\end{eqnarray}
where the integration constant $C$ is determined by the initial
condition $\varepsilon(t=0)=0$.

The creep rupture of the viscoelastic bundle can be interpreted so
that for $\sigma_{\rm 
  o} \leq \sigma_{\rm c}$ the system suffers only a partial failure
  which implies an infinite lifetime $t_f = \infty$ and the emergence
  of a macroscopic stationary state, while above the critical load
$\sigma_{\rm o} > \sigma_{\rm c}$ global failure occurs at a finite
time $t_f$, which can be determined by
evaluating the integral Eq.\ (\ref{eq:integ}) over the whole domain of
definition of 
$P(\varepsilon)$. 

Below the critical point $\sigma_{\rm 
  o} \leq \sigma_{\rm c}$ the
bundle relaxes to the stationary deformation $\varepsilon_s$ through
a decreasing breaking activity. To find 
the characteristic time scale of this relaxation process the behavior 
of $\varepsilon(t)$ has to be analyzed in the vicinity of
$\varepsilon_s$. It is useful to introduce a new variable $\delta$ as 
$\delta(t) = \varepsilon_s - \varepsilon(t)$. The governing
differential equation of $\delta$ can be obtained from Eq.\
(\ref{eq:eom}) by expanding it around $\varepsilon_s$
\begin{eqnarray}
  \label{eq:delta}
{\displaystyle   \frac{d\delta}{dt} = -\frac{E}{\beta}
  \left[1-\frac{\varepsilon_s 
      p(\varepsilon_s)}{1-P(\varepsilon_s)} \right]\delta. } 
\end{eqnarray}
The solution of Eq.\ (\ref{eq:delta}) has the form $\delta \sim
\exp{\left[-t/\tau\right]}$, where $\tau$ is the characteristic time
scale of the relaxation process
\begin{eqnarray}
  \label{eq:tau}
{\displaystyle   \tau } = \frac{\beta}{E}
  \frac{1}{\left[ 1- {\displaystyle \frac{\varepsilon_s 
      p(\varepsilon_s)}{1-P(\varepsilon_s) }} \right]}. 
\end{eqnarray}
It is a very important question how the relaxation time $\tau$ changes 
when the external driving approaches the critical point $\sigma_{\rm
  c}$ from below. Based on Eq.\ (\ref{eq:series}) it can be simply
shown that
\begin{eqnarray}
  \label{eq:tau_crit}
  \tau \sim \left(\sigma_{c} - \sigma_{\rm o}\right)^{-1/2}, \ \ \ 
  \mbox{for} \ \ \ \sigma_{o} < \sigma_{\rm c},
\end{eqnarray}
which means that approaching the critical point from below the
relaxation time of the system diverges according to a universal power
law with an exponent $-1/2$ independent on the form of disorder
distribution. Note that a similar power law divergence of the
number of successive relaxation steps was found in Refs.\
\cite{chakrab1,chakrab2} for a dry 
fiber bundle subjected to a constant load. 

Above the critical point the behavior of the lifetime of the system
can be analyzed analogously when $\sigma_{\rm o}$ goes to $\sigma_{\rm
  c}$ from above.
When $\sigma_{\rm o}$ is in the vicinity of 
$\sigma_{\rm c}$, {\it i.e.} $\sigma_{\rm o}=\sigma_{\rm c}+\Delta
\sigma_{\rm o}$, where 
$\Delta \sigma_{\rm o} << \sigma_{\rm c}$, it can be expected that the 
curve of $\varepsilon(t)$ falls very close to $\varepsilon_{\rm c}$ for a very
long time and the breaking of the system occurs suddenly.
Hence, the total time to failure,
{\it i.e.} the  
integral in Eq.\ (\ref{eq:integ}), is dominated by the 
region close to $\varepsilon_{\rm c}$ when $\Delta \sigma_{\rm o}$ is
small. Making 
use of the power series expansion Eq.\ (\ref{eq:series}) the
integral in Eq.\ (\ref{eq:integ}) can be rewritten as
\begin{eqnarray}
  \label{eq:time}
  t_f \approx  \beta \int d\varepsilon 
  \frac{1-P(\varepsilon)}{\Delta \sigma_{\rm o} - A(\varepsilon_{\rm c}-\varepsilon)^2}, 
\end{eqnarray}
which has to be evaluated over a small $\varepsilon$
interval in the vicinity of $\varepsilon_{\rm c}$. After 
performing the integration 
it follows 
\begin{eqnarray}
  \label{critical}
  t_f \approx (\sigma_{\rm o} - \sigma_{\rm c})^{-1/2}, \qquad \mbox{for} \qquad
  \sigma_{\rm o} > \sigma_{\rm c}. 
\end{eqnarray}
Thus, $t_f$ has a power law divergence at $\sigma_{\rm c}$ with
a universal exponent $-\frac{1}{2}$ independent of the
specific form of the disorder  distribution $P(\varepsilon)$,
similarly to $\tau$ below the critical point.

\subsubsection{Finite size effect}
In the above analytic treatment the size of the system, {\it i.e.} the 
number of fibers in the bundle, is infinite. However, it can be
expected that the lifetime of a finite bundle has a non-trivial size
scaling even in the case of global load sharing.
During the creep rupture process the fibers of a finite bundle break
one-by-one in the increasing order of their breaking thresholds. Let
$\varepsilon_{j}, \ j=1, \ldots , N$ denote the breaking thresholds
assigned to the fibers in a realization of the bundle. 
Since fibers break one-by-one, the actual load on the intact fibers
after the failure of $i$ fibers is $\sigma_i = \sigma_o N/(N-i)$ where
$i=0, \ldots , N-1$, and the time $\Delta
t(\varepsilon_{i},\varepsilon_{i+1})$ between the breaking of 
the $i$th and $i+1$th fibers (in the ordered series) reads as 
\begin{eqnarray}
\label{eq:t_i}
  \Delta t(\varepsilon_{i},\varepsilon_{i+1}) = -\frac{\beta}{E} \left[ \ln 
  \left(\varepsilon_{i+1}-\frac{\sigma_i}{E}\right) - \ln 
\left(\varepsilon_{i}-\frac{\sigma_i}{E}\right) \right].  
\end{eqnarray}
The lifetime $t_f$
of a sample of $N$ fibers takes the form
\begin{eqnarray}
  \label{eq:life}
  t_f(N) = \sum_{i=0}^{N-1} \Delta t(\varepsilon_{i},\varepsilon_{i+1}).
\end{eqnarray}
In order to determine how $t_f$ 
depends on $N$, Eq.\ (\ref{eq:life}) has to be averaged over many 
realizations of the disorder distribution, which can be
performed analytically. The details of the analytic calculations are
summarized in the Appendix. Finally, the average lifetime $\left<
t_f(N) \right>$ of a bundle of $N$ fibers can be cast
in the form 
\begin{eqnarray}
  \label{eq:size_scal}
  \left< t_f(N) \right> \approx t_f(\infty) + \frac{\beta \sigma_o}{N}\int
  \frac{d\varepsilon E\varepsilon 
    P(\varepsilon)\left[1-P(\varepsilon)\right]}{\left[\sigma_o-E
      \varepsilon(1-P(\varepsilon))         
    \right]^3}.
\end{eqnarray}
Eq.\ (\ref{eq:size_scal}) shows that for finite bundles
the average lifetime $\left< t_f(N)\right>$ converges to the lifetime of 
the infinite bundle $t_f(\infty)$ as $\sim 1/N$ with increasing number of
fibers $N$. It is interesting to note that in the
case of global load sharing the average strength of the bundle
$\sigma_c$ does not have a size dependence.

\subsubsection{Simulation technique}
Most of the analytic results of the previous sections, except for the
finite size scaling of $t_f$, were obtained for infinite
bundles. Computer simulations of the creep rupture of finite bundles
are needed to justify the validity of analytic predictions for finite
systems, and to be able to model the rupture process of realistic
finite systems. 
\begin{figure} 
\begin{center}
\psfrag{aa}{{\large $t_f$}}
\psfrag{bb}{{\large $t/t_o$}}
\psfrag{cc}{{\Large $\varepsilon/\varepsilon_m$}}
\psfrag{dd}{{\Large $\varepsilon_{\rm c}$}}
\epsfig{bbllx=5,bblly=0,bburx=200,bbury=175, file=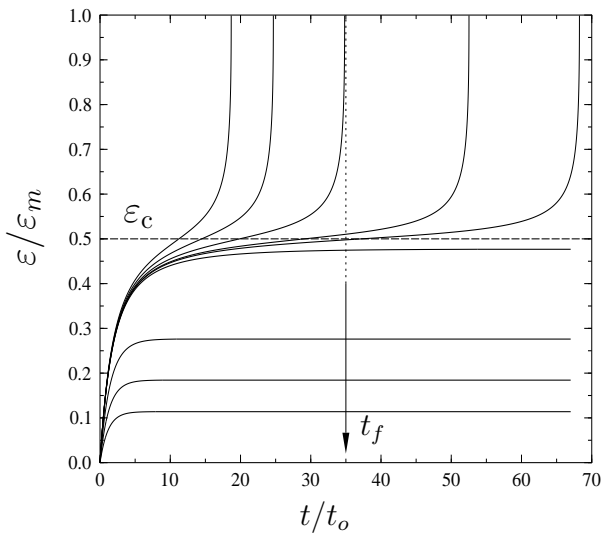, width=9cm}
\caption{$\varepsilon(t) $ 
    for several values of $\sigma_{\rm o}$ below and above
    $\sigma_{\rm c}$ for a bundle of $10^7$ fibers. The critical
    strain $\varepsilon_{\rm c}$ and the time to 
    failure $t_f$ for one example are indicated. $t_o$ denotes the
    characteristic time $t_o=\beta/E$ of the system.}
\label{fig:eps}
\end{center}
\end{figure}
In the framework of GLS an efficient simulation technique can be
worked out for the failure process.
Based on the arguments of the previous subsection, the GLS simulation
of the creep process of a bundle of $N$ fibers proceeds as follows:
$(i)$ random breaking thresholds 
$\varepsilon_{i}, \ i=1, \ldots , N$ are drawn from a
probability distribution $p$, then the thresholds 
are put into increasing order. $(ii)$ The time $\Delta t(\varepsilon_i,
\varepsilon_{i+1})$ between the breaking of 
the $i$th and $i+1$th fibers is calculated according to Eq.\
(\ref{eq:t_i}). 
$(iii)$ The time elapsed till the breaking of the $i$th fiber is
obtained as 
$t(\varepsilon_{i}) = \sum_{j=0}^{i-1} \Delta t(\varepsilon_i,
\varepsilon_{i+1})$, from which the deformation as a function of time
$\varepsilon(t)$ can be determined by inversion.
The lifetime $t_f$ of the finite bundle can be obtained
making use of Eq.\ (\ref{eq:life}). 

For simulations we considered a uniform distribution of failure
thresholds between $0$ and a maximum value $\varepsilon_m$ with the
probability density function
$p(\varepsilon)=1/\varepsilon_m$, and distribution function
$P(\varepsilon)=\varepsilon/\varepsilon_m$. In this case
the stationary solution, the critical load and the corresponding
critical strain can be obtained as 
$\sigma_{\rm o} = E\varepsilon[1-\varepsilon/\varepsilon_m]$,
$\sigma_{c}=E\varepsilon_m/4$,
$\varepsilon_{c}=\varepsilon_m/2$, respectively. 
$\varepsilon_m=1$ was set in all the simulations. 
The deformation-time diagram $\varepsilon(t)$ obtained by simulations
is presented in Fig.\ \ref{fig:eps} for several different values of
$\sigma_o$ below and above $\sigma_c$. 
The two regimes depending on the value of the
external load $\sigma_o$ can be clearly distinguished.
\begin{figure} 
\begin{center}
\epsfig{bbllx=160,bblly=378,bburx=477,bbury=668, file=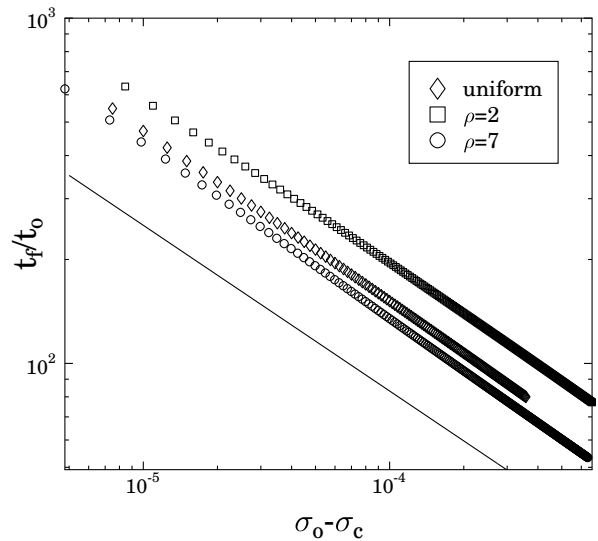,
width=8.0cm} 
\caption{The lifetime of a bundle of $10^7$ fibers as a function of
  $\sigma_o-\sigma_c$ above the 
  critical point for three different disorder distributions, {\it
    i.e.} uniform and Weibull distributions with $\rho =2, 7$ have been
  considered. The straight line of slope $-0.5$ is drawn to guide the
  eye. 
} 
\label{fig:lf_gls}
\end{center}
\end{figure} 

To test the validity of the universal power law
behavior of $t_f$ as a function of the distance from the critical
load given by Eq.\ (\ref{critical}), simulations were
performed with various disorder distributions, {\it i.e.} besides the
uniform distribution the Weibull distribution of the form
$P(\varepsilon) = 1-\exp{\left[-\left(\varepsilon/\lambda\right)^{\rho}
  \right]}$ was employed.
 The value of the characteristic strain
$\lambda$ was set to one, and the shape of the distribution was
controlled by varying the value of $\rho$. The 
results are presented in Fig.\ \ref{fig:lf_gls}, where an excellent
agreement of the simulations and the analytic results can be
observed. 
Fig.\ \ref{fig:lf_gls} supports that the exponent of $t_f$ as a
function of $\sigma_o-\sigma_c$ is universal, it does not depend on
the specific form of the disorder distribution. 

\begin{figure} 
\begin{center}
\epsfig{bbllx=127,bblly=378,bburx=477,bbury=668, file=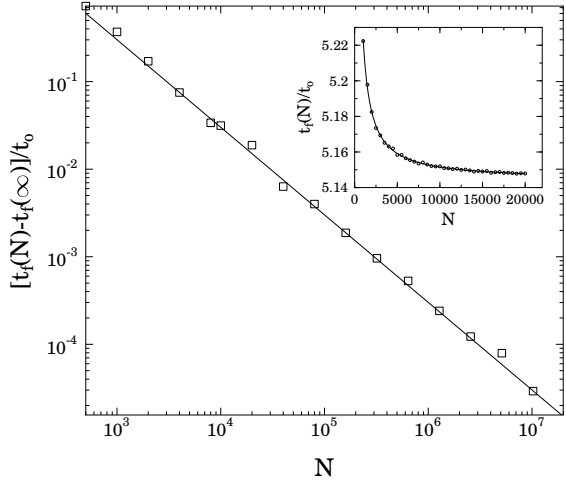,
width=8.0cm} 
\caption{The size dependent lifetime of the bundle. The inset shows
  the lifetime of relatively small systems on a linear plot, while in
  the main figure the difference of the lifetime of finite bundles and
the infinite one obtained from Eq.\ (\ref{eq:size_scal}) can be
seen on a double logarithmic plot. The slope of the fitted straight line is
$-0.99\pm0.02$.} 
\label{fig:sizescaling}
\end{center}
\end{figure}

To study the finite size scaling of the time to failure $t_f$ a uniform
distribution was used for the failure thresholds. The value of the
external load was fixed above $\sigma_c$ and the number of fibers $N$
was varied from $5\times 10^2$ to $10^7$. Averages were calculated over
$10^4$ samples for each system size $N$. The results obtained by
simulations are presented in Fig.\ \ref{fig:sizescaling}, where 
an excellent agreement of simulations and analytic
results can be observed for four orders of magnitude in the system
size $N$. 

\subsection{Slowly relaxing fibers}

\subsubsection{Analytic model}
Another important microscopic mechanism which can lead to macroscopic
creep is the slow relaxation following
fiber failure.  
In this case, the components of the solid are linearly elastic until
they break, 
however, after breaking they undergo a slow relaxation process, which
can be caused, for instance, by the sliding of broken fibers with
respect to the matrix material or by the creeeping matrix. 
To take into account this effect, our approach is based on the model
introduced in Refs.\ \cite{mcmeek,curtin}, where the response of a
viscoelastic-plastic matrix reinforced with elastic and also
viscoelastic fibers have been studied.
 
\begin{figure} 
\begin{center}
\epsfig{bbllx=0,bblly=-20,bburx=250,bbury=230, file=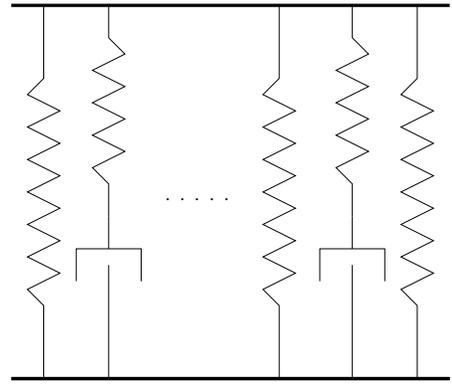, width=6cm}
\caption{The model solid when intact fibers are linearly elastic, and
  the broken ones with the surrrounding matrix are modeled by Maxwell
  elements.}  
\label{fig:maxwell}
\end{center}
\end{figure}

The model consists of $N$ parallel fibers, which break in a stress
controlled way, {\it i.e.} subjecting a bundle to a constant external
load fibers break during the time evolution of the system when the
local load on them exceeds a 
stochastically distributed breaking threshold $\sigma_i, \ \ i=1,\ldots,N$.    
Intact fibers are assumed to be linearly elastic {\it i.e.} $\sigma=E_{\rm
  f}\varepsilon_{\rm f}$ holds until they break, and hence, 
for the deformation rate it applies
\begin{eqnarray}
  \label{eq:intact}
  \dot{\varepsilon}_{\rm f} = \frac{\dot{\sigma}}{E_{\rm f}}.
\end{eqnarray}
Here $\varepsilon_f$ denotes the strain and $E_f$ is the Young modulus
of intact fibers, respectively.
The main assumption of the model is that when a fiber breaks its load
does not drop to zero instantaneously, instead  
it undergoes a slow relaxation process introducing a time scale into
the system. 
In order to capture this effect, the broken fibers with the
surrounding matrix material are modeled by
Maxwell elements as illustrated in 
Fig.\ \ref{fig:maxwell}, {\it i.e.} they
are conceived as a serial coupling of a spring and a dashpot which
results in a non-linear response 
\begin{eqnarray}
  \label{eq:broken}
  \dot{\varepsilon}_{\rm b} = \frac{\dot{\sigma}_{\rm b}}{E_{\rm b}} +
  B\sigma_{\rm b}^{\rm m},
\end{eqnarray}
where $\sigma_b$ and $\varepsilon_{\rm b}$ denote the time dependent load
and deformation of a broken fiber, respectively. The 
relaxation of the broken fiber is characterized by three parameters $E_b,
B,$ and $m$, where $E_b$ is the effective stiffness of a broken fiber,
and the exponent $m$ characterizes the strength of 
non-linearity of the element.  We study the behavior
of the system for the region $m\geq 1$.  

Assuming global load sharing for the load redistribution, the
constitutive equation describing the macroscopic elastic behavior  
of the composite reads as 
\begin{eqnarray}
  \label{eq:macro}
  \sigma_{\rm o} = \sigma(t)\left[ 1-P(\sigma(t)) \right] + \sigma_{\rm b}(t)P(\sigma(t)).
\end{eqnarray}
Eq.\ (\ref{eq:macro}) takes into account that broken fibers carry also
a certain amount of load $\sigma_b(t)$, furthermore, $P(\sigma(t))$
and $1-P(\sigma(t))$ denote the fraction 
of broken and intact fibers at time $t$, respectively \cite{kun2}. It
can be seen from Eq.\  
(\ref{eq:macro}) that under a constant external load  $\sigma_{\rm
  0}$, the load of intact fibers $\sigma$ will also be time dependent due
to the slow relaxation of the broken ones. 

Due to the boundary condition illustrated in Fig.\ \ref{fig:maxwell},
the two time derivatives have 
to be always equal
\begin{eqnarray}
  \label{eq:condit}
   \dot{\varepsilon}_{\rm f} = \dot{\varepsilon}_{\rm b}.
\end{eqnarray}
The differential
equation governing the time evolution of the system can be obtained
by expressing $\sigma_{\rm b}$ in terms of $\sigma$ from Eq.\ (\ref{eq:macro})
and substituting it into Eq.\ (\ref{eq:broken}) and finally into Eq.\
(\ref{eq:condit}) 
\begin{eqnarray}
  \label{eq:eom_max}
 && \dot{\sigma}\left\{ \frac{1}{E_{\rm f}}-\frac{1}{E_{\rm
        b}}\left[1-\frac{1}{P(\sigma)}+\frac{p(\sigma)}{P(\sigma)^2}(\sigma-\sigma_{\rm
        o})\right]\right\} = \\ \nonumber
&& B \left[\frac{\sigma_{\rm
        o}-\sigma\left[1-P(\sigma)\right]}{P(\sigma)} \right]^m.   
\end{eqnarray}
In order to determine the initial condition for the integration of
Eq.\ (\ref{eq:eom_max}) the breaking process of fibers has to be
analyzed. Subjecting the undamaged specimen to an external stress
$\sigma_o$ all the fibers attain this stress value immediatelly
due to the linear elastic response. Hence the time evolution of the
system can be obtained by integrating  Eq.\ (\ref{eq:eom_max}) with
the initial condition $\sigma(t=0) = \sigma_o$. 
Since intact fibers are linearly elastic, the deformation-time history
$\varepsilon(t)$ of the model can be 
deduced as $\varepsilon(t) = \sigma(t)/E_f$, which has an initial jump
to $\varepsilon_o = \sigma_o/E_f$. It follows that those
fibers which have breaking thresholds $\sigma_i$ smaller than the
externally imposed $\sigma_o$ immediately break.

To characterize the macroscopic behavior of the composite the solutions
$\sigma(t)$ of Eq.\ (\ref{eq:eom_max}) have to be analyzed at different
values of the external load $\sigma_o$.  
Similarly to the previous model, two different regimes of $\sigma(t)$ can
be distinguished depending on the value of $\sigma_o$: if the
external load falls below a critical value $\sigma_{\rm c}$ a
stationary solution $\sigma_s$ of the governing equation exists which can
be obtained by setting $\dot{\sigma} = 0$ in Eq.\ (\ref{eq:eom_max})
\begin{eqnarray}
  \label{eq:statio_max}
  \sigma_o = \sigma_s\left[ 1-P(\sigma_s)\right].
\end{eqnarray}
This means that until Eq.\ (\ref{eq:statio_max}) can be solved
for $\sigma_s$ the solution $\sigma(t)$ of Eq.\ (\ref{eq:eom_max}) converges
asymptotically to $\sigma_s$ resulting in an infinite lifetime
$t_f$ of the composite. Note that Eq.\ (\ref{eq:statio_max}) provides also 
the asymptotic 
constitutive behavior of the model which can be measured by
quasistatic loading. 
If the external load falls above the critical value the
deformation rate $\dot{\varepsilon} = \dot{\sigma}/E_f$ remains always 
positive resulting in a macroscopic rupture in a finite time $t_f$. It
follows from Eq.\ (\ref{eq:statio_max}) that the critical load
$\sigma_c$ of creep rupture
coincides with the static fracture strength of the composite.    

The behavior of the system shows again universal aspects in the
vicinity of the critical point. Below the critical point the
relaxation  
of $\sigma(t)$ to the stationary solution $\sigma_s$ is governed by a
differential equation of the form
\begin{eqnarray}
\frac{d\delta}{dt} \sim \delta^m,
\end{eqnarray}
where $\delta$ denotes the difference $\delta(t) = \sigma_s - \sigma(t)$. Hence,
the characteristic time scale $\tau$ of the relaxation process only emerges
if $m=1$, furthermore, in this case also $\tau \sim
(\sigma_c - \sigma_o)^{-1/2}$ holds when approaching the critical
point. However, for $m 
> 1$ the relaxation process is characterized by $\delta(t) = a
t^{1/1-m}$, where $a \rightarrow 0$ with $\sigma_0 \rightarrow
\sigma_c$. 

Similarly to the previous model, it can also be shown that the
lifetime $t_f$ of the bundle has a power law 
divergence when the external load approaches the critical point from 
above 
\begin{eqnarray}
  \label{eq:tau_max}
    t_f \sim \left(\sigma_{0} - \sigma_{\rm c}\right)^{-(m-1/2)}, \ \ \ 
  \mbox{for} \ \ \ \sigma_{o} > \sigma_{\rm c}.
\end{eqnarray}
The exponent is universal in the sense that it is independent on the
disorder distribution, however, it depends on the stress exponent $m$,
which characterizes the non-linearity of broken fibers. 

\subsubsection{Simulation technique}

Subjecting a finite bundle of $N$ fibers to $\sigma_o$ external stress
those fibers whose failure threshold falls below $\sigma_o$ break
immediately. The
number $N_o$ of intially breaking fibers can be estimated from the
disorder distribution as $N_o \approx NP(\sigma_o)$. In the presence
of broken fibers the system slows down and the remaining fibers
of the bundle break
one-by-one in the increasing order of their breaking thresholds
$\sigma_{N_o+1} < \sigma_{N_o+2} < \sigma_{N}$. In order to construct
an efficient simulation technique one has 
to determine the time elapsed between two consecutive breakings during
the creep process. 

The macroscopic constitutive equation for a system of $N$ fibers when
$i$ fibers have already failed can
be written as
\begin{eqnarray}
  \label{eq:si_macro}
  \sigma_{\rm o} = \sigma\frac{N-i}{N} + \sigma_{\rm b} \frac{i}{N}.
\end{eqnarray}
\begin{figure} 
\begin{center}
\epsfig{bbllx=150,bblly=350,bburx=490,bbury=650, file=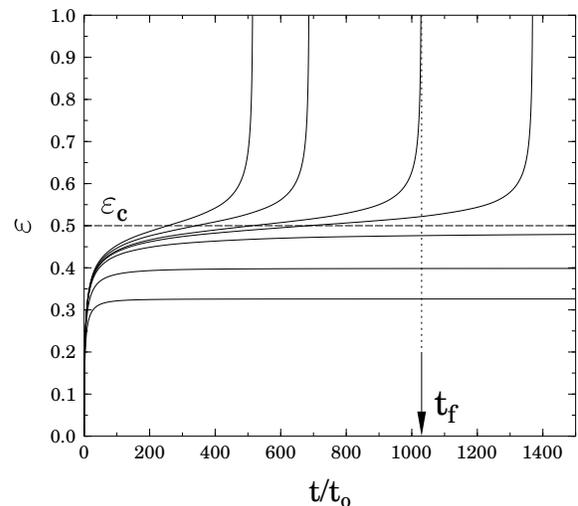,
  width=8.0cm} 
\caption{$\varepsilon$ as a function of $t$ for several value of
  $\sigma_o$ below and above $\sigma_c$. $N=10^7$ fibers were used.}  
\label{fig:eps_maxwell}
\end{center}
\end{figure} 
Making use of Eqs.\ (\ref{eq:broken},\ref{eq:condit}), the differential
equation describing the time evolution of the load of intact fibers
$\sigma$ can be cast in the form 
\begin{eqnarray}
  \label{eq:evol_max}
  \dot{\sigma}\left[\frac{1}{E_f} - \frac{1}{E_b}\left(1-\frac{N}{i} \right)
  \right] = B\left(\frac{N}{i}\right)^m f_i(\sigma)^m,
\end{eqnarray}
where $f_i(x)$ is introduced for brevity as
\begin{eqnarray}
  \label{eq:f_i}
  f_i(x) = \sigma_o - \frac{N-i}{N}x
\end{eqnarray}
The time $\Delta t(\sigma_i, \sigma_{i+1})$ elapsed between the breaking of the
$i$th and $i+1$th fiber 
can be determined by integrating Eq.\ (\ref{eq:evol_max}) from $\sigma_i$
to $\sigma_{i+1}$, which yields for $m\neq 1$
\begin{eqnarray}
  \label{eq:ti_max}
 \Delta t(\sigma_i, \sigma_{i+1})  = \frac{K_i}{(m-1)} \left[
   f_i(\sigma_{i+1})^{1-m}-f_i(\sigma_{i})^{1-m}  \right], 
\end{eqnarray}
and the multiplication factor $K_i$ reads as
\begin{eqnarray}
  \label{eq:ki_max}
  K_i = \frac{N}{N-i}\left( \frac{i}{N}\right)^m \frac{1}{B}\left[\frac{1}{E_f} -
    \frac{1}{E_b}\left(1-\frac{N}{i} \right)  \right].
\end{eqnarray}
For $m=1$ the corresponding equation has the form
\begin{eqnarray}
  \label{eq:m_eq_1}
 \Delta t(\sigma_i, \sigma_{i+1}) = K_i \left[ \ln
  f_i(\sigma_{i+1})- \ln f_i(\sigma_{i}) \right].
\end{eqnarray}

Then the simulation proceeds as in the case of viscoelastic bundles
but in the above formulas the number of broken fibers $i$ varies as
$i=N_o, N_o+1, \ldots, N-1$, so the time as a function of
$\sigma$ can be obtained as 
$t(\sigma_{i}) = \sum_{j=N_o+1}^{i} 
\Delta t(\sigma_j,\sigma_{j+1})$ from which the deformation as a
function of time $\varepsilon_{i}(t)$ can be determined, since
$\sigma=E_f\varepsilon_f$ always holds. The lifetime $t_f$ of the system
can be obtained by summing up all the $\Delta t$'s. 

\begin{figure} 
\begin{center}
\epsfig{bbllx=150,bblly=350,bburx=490,bbury=650, file=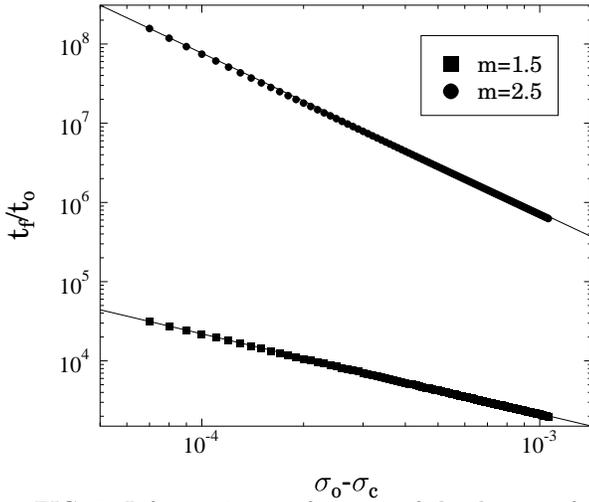,
  width=8.0cm} 
\caption{Lifetime $t_f$ as a function of the distance from the
critical point $\sigma_o -\sigma_c$ for two different values of the
parameter $m$. The number of fibers in the bundle was taken $N=10^7$.} 
\label{fig:sigma_maxwell}
\end{center}
\end{figure} 
For the purpose of explicit calculations a uniform distribution was
prescribed for the breaking thresholds $\sigma_i$ between 0 and 1. The
deformation as a function of time is plotted in Fig.\
\ref{fig:eps_maxwell} for several different values of the external
load below and above the critical load. Similarly to the previous
model the two regimes of the creeping system can be clearly
distinguished. 

To study the behavior of the time to failure as a function of the
distance from the critical point, simulations were performed for
several 
different values of the exponent $m$. In Fig.\ \ref{fig:sigma_maxwell}
the results are presented for $m=1.5$ and $m=2.5$. 
The slope of the fitted straight lines agrees very well with the
analytic predictions of Eq.\ (\ref{eq:tau_max}).

\begin{figure} 
\begin{center}
\epsfig{bbllx=150,bblly=350,bburx=490,bbury=650, file=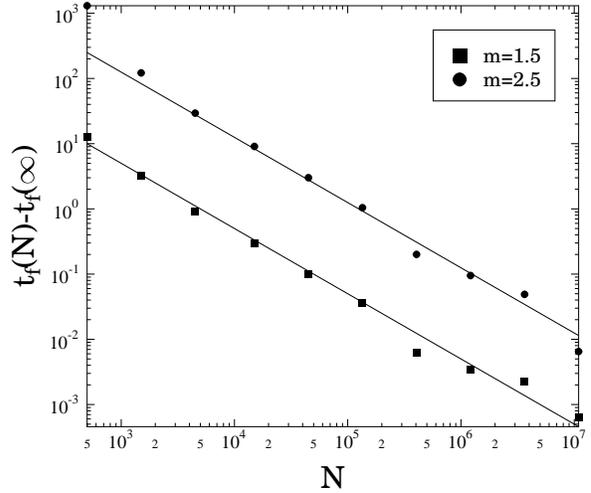,
  width=8.0cm} 
\caption{Size dependence of the lifetime $t_f$ for two different
values of the parameter $m$. } 
\label{fig:size_maxwell}
\end{center}
\end{figure}

The size scaling of the time to failure $t_f$ was analyzed by
simulating the creep rupture of bundles of size $N=5\cdot 10^2 - 10^7$
setting a uniform distribution for the breaking thresholds. We
found that $t_f(N)$ converges to the lifetime of the infinite system
$t_f(\infty)$ according to the universal law Eq.\ (\ref{eq:size_scal})
independently on the value of the exponent $m$. In Fig.\
\ref{fig:size_maxwell} the best fit was obtained for both curves with
slope $-1.0\pm 0.05$ for both $m$ values.  
\section{Conclusions}
The creep rupture of fibrous materials occurring under a
steady external load is microscopically a rather complex phenomenon
depending on a diversity 
of possible material specific mechanisms. Therefore, on the one hand,
it is impossible to work out a general theoretical framework which
takes into account all the features of the process and has predictive
power, and on the other hand,
it is very important to reveal universal aspects of the creep process,
which do not depend on specific material properties relevant at the
micro level. 

In the present paper we studied the creep rupture of
fibrous materials in the framework of 
fiber bundle models taking into account two possible microscopic
mechanisms of creep: $(i)$ in the first approach the fibers 
themselves are viscoelastic and they break when their
deformation exceeds a stochastically distributed threshold value,
$(ii)$ in the second model
the fibers are linearly elastic until they break,
however, after breaking their relaxation is not instantaneous but  
the creeping  matrix 
introduces an intrinsic time scale for the relaxation.
The first model can be relevant for natural fiber composites like wood
which are composed of viscoelastic fibers
\cite{laufenberg,pu,acoust,gerhards}, while the second model can 
provide an adequate description of metal matrix composites reinforced
by brittle fibers \cite{weber}. Analytical and numerical calculations showed in
both models that increasing the
external load on a specimen a transition takes place from a
partially failed state of infinite lifetime to a state where global
failure occurs at a finite time. The critical load turned to be the
static fracture strength of the material. It was found that although
the details of the two model systems are different,
in the vicinity of the critical point they exhibit universal
behaviour which can be relevant for experiments. 

\section{Acknowledgment}
This work was supported by the project SFB381, by the NATO grant
PST.CLG.977311 and by OTKA T037212.
F.\ Kun  acknowledges financial support of
the B\'olyai J\'anos Foundation of
the Hungarian Academy of Sciences and of the Research Contract FKFP
0118/2001.

\section{Appendix}

Here we provide the derivation of the average lifetime for the
general case when the lifetime  $t_f$ of a bundle with a specific
realization of the disorder can be cast in the form
\begin{eqnarray}
  \label{eq:tf_general}
  t_f = \sum_{i=0}^{N-1} \left[G(\frac{i}{N},x_{i+1})
    -G(\frac{i}{N},x_i) \right],
\end{eqnarray}
{\it i.e.} $t_f$ is a sum of terms which depend on the number of
broken fibers $i$ and on a single breaking threshold $x_i$ that can be
given as strain or stress. The $x_i$-s are obtained by choosing $N$ breaking
thresholds independently from a cumulative probability distribution
$P(x)$ and putting them into increasing order.
This treatment includes both models discussed in the present paper. 
The expectation value of a function $f(x_i)$ can
be determined as 
\begin{eqnarray}
  \label{eq:expect}
 \left<f(x_i) \right> &=& \int             \nonumber
  \frac{N!}{(i-1)!(N-i)!}P(x)^{i-1}\left[1- 
  P(x)\right]^{N-i} \\ && \times p(x)f(x)dx. 
\end{eqnarray}
The probability distribution in Eq.\
(\ref{eq:expect}) that the value of the $i$th largest
breaking threshold falls between $x$ and $x+dx$ 
has a sharp peak for large $N$ values for each $i$. The above integration can be 
carried out by expanding the distribution about its peak. After
expanding the result in terms of $1/N$ and neglecting higher order
terms we arrive at
\begin{eqnarray}
  \label{eq:aver}
  \left< f(x_i) \right> &=&  f(\overline{x}_i)   \nonumber
  +\frac{1}{2N}P(\overline{x}_i)(1-P(\overline{x}_i)) \\
 && \times \left[
    \frac{f^{''}(\overline{x}_i)}{[P^{'}(\overline{x}_i)]^2}-
    \frac{f^{'}(\overline{x}_i)P^{''}(\overline{x}_i)}{[P^{'}
      (\overline{x}_i)]^3}\right],   
\end{eqnarray}
where $\overline{x}_i$ is defined implicitly by
$P(\overline{x}_i)=i/(N+1)$. Applying Eq.\ (\ref{eq:aver}) to Eq.\
(\ref{eq:tf_general}) the resulting summation can be approximated 
by integrals replacing $i/N$ by the equivalent
$P(\overline{x}_i)(1+1/N)$. Neglecting corrections higher order in
$1/N$ after straightforward calculations we arrive at
\begin{eqnarray}
  \label{eq:atlagos}
 && \left< t_f \right> \approx \int dx \partial_2     \nonumber
  G(P(x),x) \\ && +\frac{1}{2N}\int dx
  P(x)\left(1-P(x)\right)\partial_1^2\partial_2G(P(x),x), 
\end{eqnarray}
where $\partial_2G(y,x)\equiv \partial G(y,x)/\partial x$, and
$\partial_1^2\partial_2 G(y,x) \equiv \partial^3 G(y,x)/\partial^2y\partial
x$.  
Substituting the actual form of $G(y,x)$ for a specific model the
complete form of the size scaling of lifetime can be
obtained. However, it can be seen in the general expression Eq.\
(\ref{eq:atlagos}) that the first term provides the lifetime of the
infinite bundle and the only size dependence is in the prefactor of
the second term. Eq.\
(\ref{eq:atlagos}) states that if the lifetime can be written in the
form of Eq.\ (\ref{eq:tf_general}) the lifetime of finite bundles
converges to that of the infinite one as $1/N$ with increasing number
of fibers $N$.

\end{multicols}

\end{document}